# Neural network potential molecular dynamics simulations of (La,Ce,Pr,Nd)$_{0.95}$(Mg,Zn,Pb,Cd,Ca,Sr,Ba)$_{0.05}$F$_{2.95}$

**Yoyo Hinuma***

Department of Energy and Environment, National Institute of Advanced Industrial Science and Technology (AIST), 1-8-31, Midorigaoka, Ikeda, Osaka 563-8577, Japan

*y.hinuma@aist.go.jp

**Abstract**

Tysonite structure fluorides doped with divalent cations, represented by Ce$_{0.95}$Ca$_{0.05}$F$_{2.95}$, are a class of good F$^-$ ion conductors together with fluorite-structured compounds. Computational understanding of the F$^-$ conduction process is difficult because of the complicated interactions between three symmetrically distinct F sites and the experimentally observed change in the F diffusion mechanism slightly above room temperature, effectively making first principles molecular dynamics (FP-MD) simulations, which are often conducted well above the transition temperature, useless when analyzing behavior below the transition point. Neural network potential (NNP) MD simulations showed that the F diffusion coefficient is higher when the divalent dopant cation size is similar to the trivalent cation size. The diffusion behavior of F in different sites changes at roughly 500 K in Ce$_{0.95}$Ca$_{0.05}$F$_{2.95}$ because only the F1 site sublattice contributes to F diffusion below this temperature but the remaining F2 and F3 sublattices becomes gradually active above this temperature. The paradox of higher diffusion coefficients in CeF$_3$-based compounds than similar LaF$_3$-based compounds even though the lattice parameters are larger in the latter may be caused by a shallower potential of Ce and F in CeF$_3$ compared to the LaF$_3$ counterparts.

## 1. Introduction

Solid state ionic conductors have many applications including batteries, solid-oxide fuel cells, electrochemical sensors, electrochromic windows, and oxygen-separation membranes. [1] F$^-$ is a well-studied conductor; experimental conductivities of various F$^-$ ion conductors are summarized in Motohashi et al. [2] and Zhang et al.[3] A computational high-throughput search with a decoupled, dynamic, and iterative scheme was used to find intercalation F$^-$ conductors. Their nudged elastic band (NEB) calculations suggest that the activation barrier for F- diffusion decreases when F- is closer to cations and more distant from anions at the transition state and the diffusion pathway is shorter than 3.8 Å, which excludes F⁻ hops to a second nearest-neighbor fluoride vacancy. [4] Crystal structures that often allow high F⁻ conductivity are tysonite, fluorite, PbSnF$_4$ analogues, and perovskite. [3] A prototypical tysonite structure conductor is divalent cation doped LaF$_3$, which is considered for solid electrolytes of F⁻ ion batteries [5-6] and F⁻ ion sensors.[7] What cations that can form the tysonite structure has been investigated in detail; the $R$F$_3$ systems and phase diagrams of $R$F$_3$-$R'$F$_3$ systems, where $R$ and $R'$ are either Y or lanthanides La-Lu, have been studied intensively[8], where part of the motivation is identifying negative thermal expansion from a phase transition-type mechanism that is found in some of these systems.[9]

Doping of divalent cations can enhance the F⁻ conductivity in tysonite structure fluorides. The F⁻ conductivity of LaF$_3$ and isostructural CeF$_3$ increases when a few % of La$^{3+}$ (or Ce$^{3+}$) is substituted by a divalent cation and the charge is compensated by F vacancies. [2] Synthesis of La$_{1-x}$Ba$_x$F$_{3-x}$ benefits from the similar vapor pressures of BaF$_2$ and LaF$_3$.[10-11] The redox potential of La/LaF$_3$ is too low for a battery anode, at -2.4V versus Pb/PbF$_3$, though In+La$_{0.9}$Ba$_{0.1}$F$_{2.9}$ has a redox potential of -1.8 eV that prevents electrolyte decomposition.[12] On the other hand, the large potential window of LaF$_3$ makes it suitable for the solid electrolyte of an all solid state F⁻ battery.

$^{19}$F nuclear magnetic resonance (NMR) experiments consistently suggest that F⁻ diffusion in tysonite structure fluorides at room temperature happens between F1 sites, and increasing temperature slightly above room temperature makes possible ion exchange between F1 and F2 or F3 sublattices. A study on La$_{0.99}$Sr$_{0.01}$F$_{2.99}$ finds that F3 atoms start moving at 336K and becomes fully mobile at 471K, and F2 starts moving at this temperature and is fully mobile at 539 K [13]. Another study of La$_{0.95}$Sr$_{0.05}$F$_{2.99}$ reports exchange between F1 and F2 or F3 at above ~330 K and exchange between all structural positions above ~430 K[14]. Ref. [15] suggests that, in CeF$_3$, exchange between F1 and F2 sites starts at 290±10 K and completes at 320 K, and a quasi-total exchange between F1, F2, and F3 sites is attained at 460 K. The exchanges between the three sublattices start at a lower temperature than the completion of F1 and F2 site exchange in Ce$_{1-y}$Cd$_y$F$_{3-y}$ when $y$=0.03, 0.05, and 0.07. The F2 atoms in Ce$_{0.99}$Sr$_x$F3 ($x$=0.01, 0.025, and 0.05) become mobile by 337 K[16]. NMR spectrum analysis finds that F2-F3 chemical exchange requires two successive steps via a F1 site. [17]

There are many computational studies on F diffusion in fluorite-structure fluorides. Radiation damage in CaF$_2$ was computationally evaluated by obtaining the threshold displacement energy, the minimum kinetic energy needed to permanently displace an atom in a solid, in $CaF_2$ using molecular dynamics (MD) with pair potentials and simulated statistical analysis.[18] First principles (FP) MD simulations of β-$PbF_2$ at 1000 K showed excellent agreement with experiments regarding the self-diffusion coefficient and ionic conductivity.[19] Ogawa et al. suggests, using first-principles calculations and on-the-fly machine-learning force fields, that mixing Ba and Ca in the fluorite structure (Ba,Ca)$F_2$ system reduces the F-ion Frenkel pair defect formation energy rather than the F migration barrier.[20] In a related system, FP-MD simulation on interstitial F diffusion of $LaO_{1-x}F_{1+2x}$ were conducted using different Bravais lattices for the host structure.[21]

This paper used neural network potential (NNP) MD to simulate F diffusion in (La,Ce,Pr,Nd)$_{0.95}$(Mg,Zn,Pb,Cd,Ca,Sr,Ba)$_{0.05}$F$_{2.95}$. The differences in the diffusion coefficient with the choice of trivalent and divalent cation were studied. The change in the F diffusion mechanism with temperature was clarified, together with some insight on why Ce systems have F higher conductivity than La systems even though $CeF_3$ has smaller lattice parameters than $LaF_3$.

## 2. Methods

$La_{1-x}Ba_xF_{3-x}$ and related tysonite-structure doped fluoride compounds are a computational scientist's nightmare. NEB calculations[22] finds the activation barrier by optimizing a number of intermediate images simultaneously along the reaction path. Crystal structure analysis suggests seven F diffusion paths, namely three F1-F1 paths, three F1-F2 paths, and a F1-F3 path.[23] Understanding the diffusion of F requires considering the competition between these paths. On the other hand, FP-MD simulations do not require explicit identification of diffusion paths, but its effectiveness is very limited. FP-MD is often conducted at very high temperatures on the order of ~1000 K to observe a sufficient number of atom hops, which is necessary for statistical handing, in a very short simulation time of at most ~100 ps. However, the F diffusion behavior changes at temperatures much lower than standard FP-MD temperatures.

Therefore, MD simulations were conducted using the commercially available Matlantis package from Preferred Networks with their universal PreFerred Potential (PFP) [24] version 6.0.0, a NNP trained on the Perdew-Burke-Ernzerhof (PBE) generalized gradient approximation (GGA) to density functional theory (DFT)[25]. The canonical, or constant number of atoms, volume, and temperature (NVT), ensemble, and a Nosé–Hoover thermostat[26-27] were used. The time step was 2 fs, and the simulation duration was 1,000,000 steps (2 ns).

The PFP is already fully trained by the developers based on calculations using the VASP code[28-29] and is available as a "take it or leave it" potential; the user cannot modify it. PFP v6.0.0 supports only La, Ce, Pr, Nd, Sm, and Gd among the lanthanides, which is sufficient for the calculations in this study. Handling of rare earth species that are not closed shell is very difficult in DFT, including GGA. For example, the "extra electron" in $Ce^{3+}$ compared to closed shell $Ce^{4+}$ can enter $d$ or $f$ states, and the resulting state can be guided by, for example, adjusting the Hubbard $U$. Verifying what DFT implementation is how much reasonable in what structural or physical properties is a complicated task and contains many ambiguities. Fortunately, or unfortunately, there is only one choice in PFP and there is no room for adjustment. This paper aims to verify the PFP results by comparing diffusion properties with experimental values, thereby avoiding the problem of verification using DFT results.

Fig. 1 shows the crystal structure of $LaF_3$. The space group is $P\bar{3}c1$ and its NNP-relaxed lattice parameters are $a$=7.243 Å and $c$=7.400 Å for $LaF_3$, $a$=7.205 Å and $c$=7.335 Å for $CeF_3$, $a$=7.157 Å and $c$=7.317 Å for $PrF_3$, and $a$=7.106 Å and $c$=7.280 Å for $NdF_3$, respectively. La and F are nine- and three-fold coordinated, respectively. There are three types of F sites, F1, F2, and F3, with Wyckoff positions 12$g$, 4$d$, and 2$a$, [23] and representative coordinates are (0.6908, 0.0614, 0.5796), (1/3, 2/3, 0.1848), and (0, 0, 1/4), respectively. The Wyckoff position and representative coordinates of La are 12$g$ and (0.6570, 0, 1/4), respectively. XRD and neutron diffraction refinement results for $La_{1-y}Sr_yF_{3-y}$ are reported in Ref. [30], while $CeF_3$ and $Ce_{2.95}(Ca, Sr, Ba)_{0.05}F_{2.95}$ are discussed in Refs. [16, 23].

An orthorhombic supercell of 216 formula units was used in the simulations, with lattice parameters $a$=21.614 Å, $b$=24.958 Å, and $c$=22.005 Å for $LaF_3$. Cation substitution to a divalent cation, namely Mg, Zn, Pb, Cd, Ca, Sr, or Ba, of 11 sites among the 216 cation sites were compensated by formation of 11 vacancies in F1 sites, resulting in a stoichiometry of $La_{0.95}Sr_{0.05}F_{2.95}$, for example. The initial positions of these defects were randomly generated once and the same positions were used in all calculations.

The tracer diffusion coefficient in the $x$ direction, $D_x$, was obtained by plotting the mean square displacement (MSD) in the $x$ direction, $\langle \Delta x_{in}^2 \rangle_i$, versus $n\tau$ and fitting to $\langle \Delta x_{in}^2 \rangle_i = 6D_x n\tau$. Here, $\Delta x_{in}$ is the displacement of atom $i$ in the $x$-direction at step $n$ from the initial position at $n=0$, $\langle \cdot \rangle_i$ indicates taking the mean of all relevant atoms $i$, and $\tau$ is the time step. The value of $\langle \Delta x_{in}^2 \rangle_i$ depends on $n$. Diffusion coefficients along the $y$ and $z$ directions, $D_y$ and $D_z$, respectively, were obtained similarly, and the 3D diffusion coefficient is $D=D_x+D_y+D_z$. MSDs were obtained from 1000 snapshots taken every 1000 steps. The tracer diffusion coefficient averaged over all F atoms, F1 only, F2 only, and F3 only are denoted as $D$, $D_1$, $D_2$, and $D_3$ in this paper, respectively.

The mean, over atoms, of the variance of the displacement of each atom with regard to the equilibrium point, $\langle \Delta r^2 \rangle = \langle Var_n(\Delta r_{in}) \rangle_i = \langle Var_n(\Delta x_{in}) + Var_n(\Delta y_{in}) + Var_n(\Delta z_{in}) \rangle_i$ was obtained for $La_{216}F_{648}$, $Ce_{216}F_{648}$, $Pr_{216}F_{648}$, and $Nd_{216}F_{648}$ compounds without any defects. Here, $\Delta x_{in}$, $\Delta y_{in}$, and $\Delta z_{in}$ are the displacements from the equilibrium point in the $x$, $y$, $z$ directions, respectively, of atom $i$ at step $n$, and $\Delta r_{in}^2 = \Delta x_{in}^2 + \Delta y_{in}^2 + \Delta z_{in}^2$. The variances are calculated over all $n$ for each atom $i$, which is stressed with the subscript $n$ after $Var$. This quantity describes the spatial distribution of non-diffusing atoms around the equilibrium point and values calculated over different temperatures can be used to check the harmonicity of the atom. 1000 snapshots taken every 100 steps were used for this purpose.

## 3. Results and discussion
### 3.1 $D$ dependence on stoichiometry
Fig. 2 shows the diffusion coefficients of $(La,Ce,Pr,Nd)_{0.95}(Mg,Zn,Pb,Cd,Ca,Sr,Ba)_{0.05}F_{2.95}$ at temperature $T=500$ K ($1/T=0.002$ K$^{-1}$). The vertical axis is $D_1$, $D_2$, or $D_3$, and the horizontal axis is the cation size for eight-fold coordination according to Shannon [31]. The cations are nine-folded coordinated in the tysonite structure, but the ionic radius for nine-fold coordination was missing in some systems. The nine-fold coordination radius is consistently 0.05-0.06 Å larger than the eight-folded coordination radius in relevant elements with available data[31], hence the eight-fold coordination radius was used instead.

The $D_1$ (Fig. 2(a)) is roughly an order of magnitude larger than $D_2$ and $D_3$ (Figs. 2(b,c), respectively), thus F1 is the dominantly diffusing species at this temperature. A volcano-type clear trend exists where $D_1$ is large when the divalent and trivalent cation sizes are close to each other; the ionic radius of divalent cations are shown with dashed lines and labels, and those of trivalent cations are plotted with thick dashed lines within in the range 1.11 to 1.16 Å. When the divalent cation is fixed, $D_1$ is largest and second largest when the trivalent cation is Ce and La, respectively. This may be surprising because the lattice parameter of $LaF_3$ is larger than $CeF_3$ and the activation barrier is generally lower when the lattice parameter is larger because of a wider bottleneck. The values of $D_2$ and $D_3$ are not discussed further as the values are too low and there is much margin of error.

Divalent cations Pb, Cd, Ca, and Sr are considered hereon because of the large $D_1$ compared to the remaining divalent cations.

The effect of the lattice parameter on $D_1$ is investigated in detail. Fig. 3(a) shows the $D_1$ of "La-systems", or $La_{0.95}(Pb,Cd,Sr,Ca)_{0.05}F_{2.95}$, when the lattice parameters are fixed to those of $LaF_3$, $CeF_3$, $PrF_3$, or $NdF_3$ (black circles, red upward pointing triangles, green squares, and blue downward pointing triangles, respectively) The divalent cations in the horizontal axis of Fig. 3 are in increasing order of $D$ when La, Ce, or Pr is the trivalent cation (see Fig. 2(a)). The $D$ is larger when the lattice parameter of the system is larger, as expected, regardless of the divalent cation choice. The same trend is found in Ce-, Pr-, and Nd-systems, or $Ce_{0.95}(Pb,Cd,Sr,Ca)_{0.05}F_{2.95}$, $Pr_{0.95}(Pb,Cd,Sr,Ca)_{0.05}F_{2.95}$, and $Nd_{0.95}(Pb,Cd,Sr,Ca)_{0.05}F_{2.95}$, respectively (Figs. 3(b-d), respectively).

**3.2 *D* dependence on temperature**
Fig. 4 shows the Arrhenius plot of $Ce_{0.95}Ca_{0.05}F_{2.95}$, the champion system with highest $D$ in Fig. 2(a), over all F, F1 only, F2 only, and F3 only. $D_2$ and $D_3$ increases at a much higher pace than $D_1$ when the temperature is increased above 500 K ($1/T=0.002$ K$^{-1}$) and becomes almost the same at 833 K ($1/T=0.0012$ K$^{-1}$).

These $D$ values are calculated based on the initial position of F atoms. For example, the diffusion behavior of an initially F2 atom that migrated to an F1 site and then to an F3 site is captured in $D_2$ only regardless of where it moved to. The diffusion coefficients and activation barriers of $D_1$, $D_2$ and $D_3$ should become the same in a sufficiently long simulation when F atoms can move freely between F1, F2, and F3 sites. The change in the activation barrier may partly reflect the temperature-dependent change in the prefactor instead of a gradual change in the true activation barrier. An increase in former F2 and F3 atoms that migrate to F1 sites that can diffuse easily effectively results in a prefactor that increases with temperature for originally F2 and F3 atoms, resulting in an apparently steeper Arrhenius plot with apparently larger activation barrier. On the other hand, migration of F1 atoms to F2 and F3 sites apparently flattens the Arrhenius plot and decreases the apparent activation barrier for originally F1 atoms. The Arrhenius plot of $D_1$ appears to bend near 500 K ($1/T=0.002$ K$^{-1}$), reflecting the rise of F2 and F3 diffusion above this temperature. This difference in the diffusion behavior of F1 compared to F2 and F3 is consistent with experimental findings, implying the extent of validity of the PFP regarding diffusion energetics on the studied systems.

Fig. 5 shows the Arrhenius plots of $(La,Ce)_{0.95}(Pb,Cd,Sr,Ca)_{0.05}F_{2.95}$ for $D_1$ at $1/T=0.002$ K$^{-1}$ and below and their activation barriers. The activation barriers are within the range 0.14-0.23 eV. Using Pb slightly increases the activation barrier and Ce-based compounds have a lower activation barrier than La-based compounds.

Most of the experimental activation barriers are much larger than the values in Fig. 5. The F diffusion activation barrier between F1 sites were obtained as 0.20±0.02 eV for $La_{1-x}Sr_xFe_{3-x}$ for $x \leq 0.03$, and a lower barrier for larger $x$ based on further narrowing of the residual F$_I$ dipolar linewidth with rising temperature in $^{19}$F NMR spectra[32]. The conductivities of $Ce_{0.95}Ca_{0.05}F_{2.95}$, $Ce_{0.95}Sr_{0.05}F_{2.95}$, and $Ce_{0.95}Ba_{0.05}F_{2.95}$ decrease in this order, and the activation barriers are 0.41, 0.44, and 0.45 eV. [23] The $La_{0.95}Sr_{0.05}F_{2.95}$ activation barrier is 0.50 eV[33], and that of $La_{0.95}Ba_{0.95}F_{2.95}$ from the wet chemical method is 0.47 eV, and becomes 0.42 eV after sintering at 20 hours [34]. Ref. [35] states that the single crystal $La_{0.95}Sr_{0.05}F_{2.95}$ Arrhenius plot of conductivity bends at 430K. Arrhenius plots of LaF$_3$ and slightly doped $La_{0.95}Sr_{0.05}F_{2.95}$ can bend, implying the existence of multiple diffusion mechanisms, and the bending temperature depends on how the material was prepared [36].

The apparent activation barrier based on conductivity may reflect an activation barrier other than the intrinsic activation barrier of bulk. Moreover, in a system where the conductivity $\sigma=qND/k_BT$ ($q$ is charge, $k_B$ is the Boltzmann constant, and $T$ is temperature) may be written as $N=N_0\exp(-E_{def}/k_BT)$ ($N_0$ is a constant and $E_{def}$ is a defect formation energy) and $D=D_0\exp(-E_{mig}/k_BT)$ ($D_0$ is a constant and $E_{mig}$ is the ion migration energy), the apparent activation barrier is $E_{def}+E_{mig}$. [20] This being said, large differences in the bulk activation barriers between between F1 sites only and those involving F2 or F3 sites should lead to different critical temperatures for the onset of F2 and F3 diffusion, and this match between the lowest reported experimental activation barrier and the computed barrier suggests the validity of the PFP.

### 3.3 Anharmonicity of F

This section discusses the mean, over atoms, of the variance of the displacement of each atom with regard to the equilibrium point. A very common MD analysis is to plot the mean square displacement (MSD) versus simulation time and use the slope to obtain the diffusion coefficient. In non-diffusing species, the MSD becomes a constant value regardless of simulation time. The mean, over atoms, of the variance of the displacement of each atom with regard to the equilibrium point, $\langle \Delta x^2 \rangle = \langle Var_n(\Delta x_{in}) \rangle_i$, is determined solely by the potential. Here, $\Delta x_{in}$ is the displacements in the $x$ coordinate of atom $i$ at step $n$, and $\Delta r_{in}^2 = \Delta x_{in}^2 + \Delta y_{in}^2 + \Delta z_{in}^2$. The variances are calculated over all $n$ for each atom $i$ to be averaged, which is stressed with the subscript $n$ after $Var$. The equilibrium point for each atom does not need to be obtained explicitly.

Fig. 6 shows $\langle \Delta x^2 \rangle$ and their $y$ and $z$ coordinate versions, $\langle \Delta y^2 \rangle$ and $\langle \Delta z^2 \rangle$, for Ce, F1, F2, and F3 sites of $CeF_3$. A total of 1000 snapshots were taken every 100 steps in MD simulations with 100,000 steps with 2 fs per step. There are no F vacancy sites, so F diffusion is not expected. All of $\langle \Delta x^2 \rangle$, $\langle \Delta y^2 \rangle$, $\langle \Delta z^2 \rangle$ below 250K are on a linear trend that passes through the origin, which means that the potential is harmonic.[37] $\langle \Delta x^2 \rangle$ and $\langle \Delta y^2 \rangle$ should be almost the same because of symmetry. However, $\langle \Delta z^2 \rangle$ is smaller than these in Ce and F1 sites and larger in F2 sites. A large $\langle \Delta z^2 \rangle$ means that the potential is shallow in the $z$ direction and therefore atoms can move farther from the equilibrium point.

The above observation is consistent with experimental results. The anisotropic atomic displacement factors, $U_{11}$, $U_{22}$, and $U_{33}$, from neutron diffraction of $CeF_3$ are ($U_{11}$, $U_{22}$, $U_{33}$) = (0.005, 0.003, 0.004), (0.023, 0.014, 0.005), (0.005, 0.005, 0.018) and (0.007, 0.007, 0.042) for Ce, F1, F2, and F3, respectively.[23] The order of $U_{11}$, $U_{22}$, and $U_{33}$ and $\langle \Delta x^2 \rangle$, $\langle \Delta y^2 \rangle$, of $\langle \Delta z^2 \rangle$ share the same trend.

The $\langle \Delta x^2 \rangle$ of F1 sites and $\langle \Delta z^2 \rangle$ of F2 and F3 sites break away from the harmonic trend above 300 K (see Fig. 6). Escaping the harmonic potential could lead to a higher diffusion coefficient. From the other point of view, there may be a transition temperature at 250~300 K where F diffusion abruptly becomes smaller below this temperature, though this phenomenon is not evident in the simulation of $Ce_{0.95}Ca_{0.05}F_{2.95}$ in Fig. 4. The very large anisotropy on how F atoms can move around the equilibrium point could result in very complicated diffusion paths between F sites.

Fig. 7 compares $<\Delta r^2>=<\Delta x^2>+<\Delta y^2>+<\Delta z^2>$ versus $T$ for the cation, F1, F2, and F3 atoms between $LaF_3$, $CeF_3$, $PrF_3$, and $NdF_3$. All F points are shown for reference in Figs. 7(e-g), but what is important is the harmonic regime in Figs. 7(a-d). The points for Ce are consistently slightly larger than La, while Pr and Nd points are often almost the same and typically smaller than La. This means that Ce and F atoms in $CeF_3$ are in a shallower potential compared to La and F in $LaF_3$. This shallow potential, which implies a weaker bound from the equilibrium position, of $CeF_3$ could be one cause why $Ce_{0.95}(Mg,Zn,Pb,Cd,Ca,Sr,Ba)_{0.05}F_{2.95}$ has a higher F diffusion coefficient compared to the La counterparts even though $CeF_3$ has smaller lattice parameters than $LaF_3$.

## 4. Conclusions

The diffusion coefficients of F, $D$, in the F conductor $(La,Ce,Pr,Nd)_{0.95}(Mg,Zn,Pb,Cd,Ca,Sr,Ba)_{0.05}F_{2.95}$ were obtained using NNP-MD simulations. $D$ is highest when the trivalent cation is Ce, followed by La, Pr, and Nd. A volcano plot type relation exists where divalent cations with ionic radii closer to the trivalent cation have larger $D$. Simply increasing the lattice parameters increases $D$. The diffusion process mainly happens in F1 atoms below 500 K, but the diffusion coefficients of F2 and F3 atoms increase above this temperature and becomes on par with F1 atoms above 800 K. The calculated activation barriers of $(La,Ce)_{0.95}(Pb,Cd,Ca,Sr)_{0.05}F_{2.95}$ are 0.14-0.23 eV, less than half of most, but not all, experimental reports. One reason for the higher $D$ of Ce-based systems with smaller lattice parameters than La-based systems may be the shallower potential of Ce and F in $CeF_3$ compared to the $LaF_3$ counterparts.

## Acknowledgements

The VESTA code [38] was used to draw Figure 1. This paper is based on results obtained from a project, JPNP21016, commissioned by the New Energy and Industrial Technology Development Organization (NEDO).

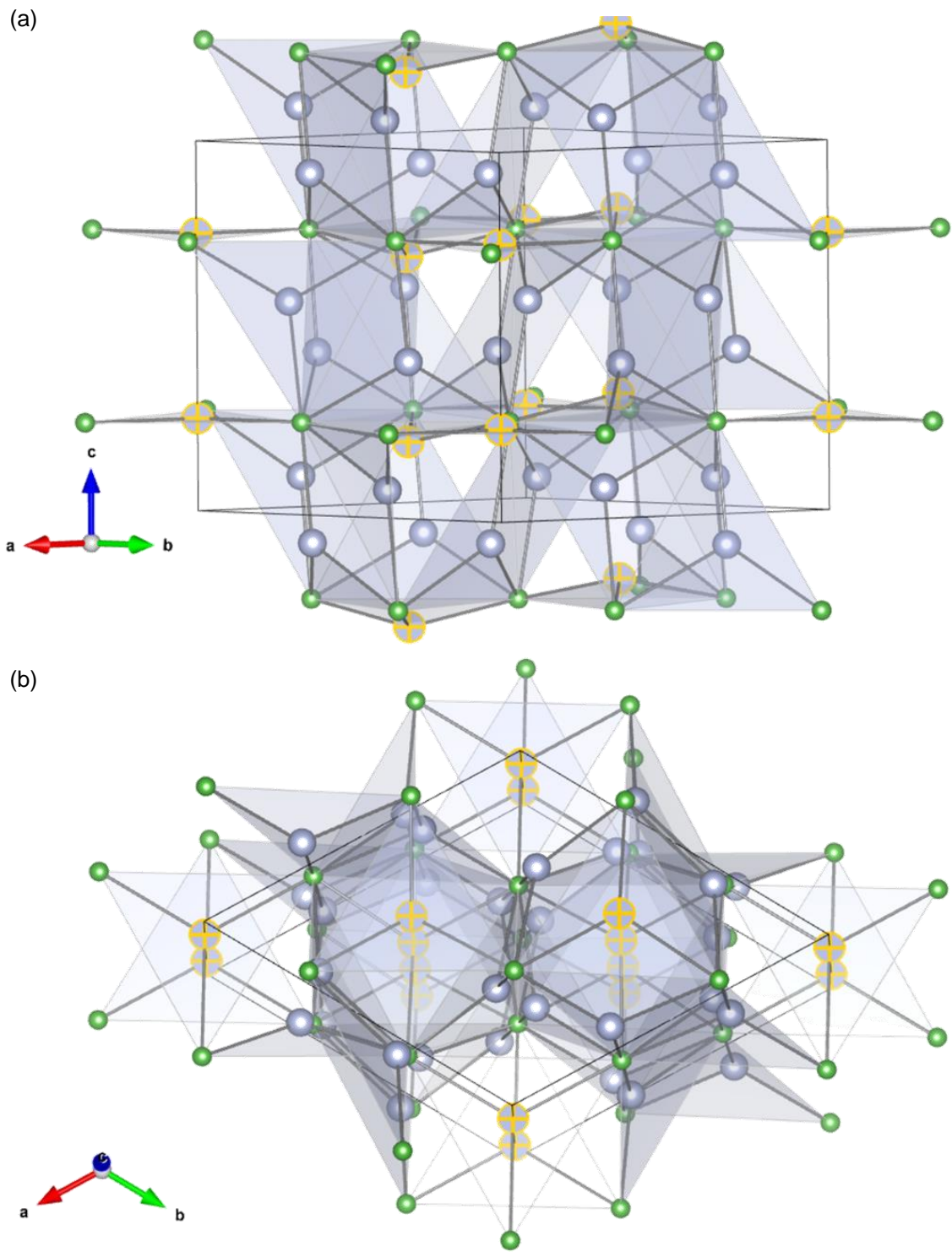

Fig. 1. CeF$_3$ viewed from two directions. Green and gray balls indicate La and F atoms, respectively. F2 and F3 atoms are shown with crosses.

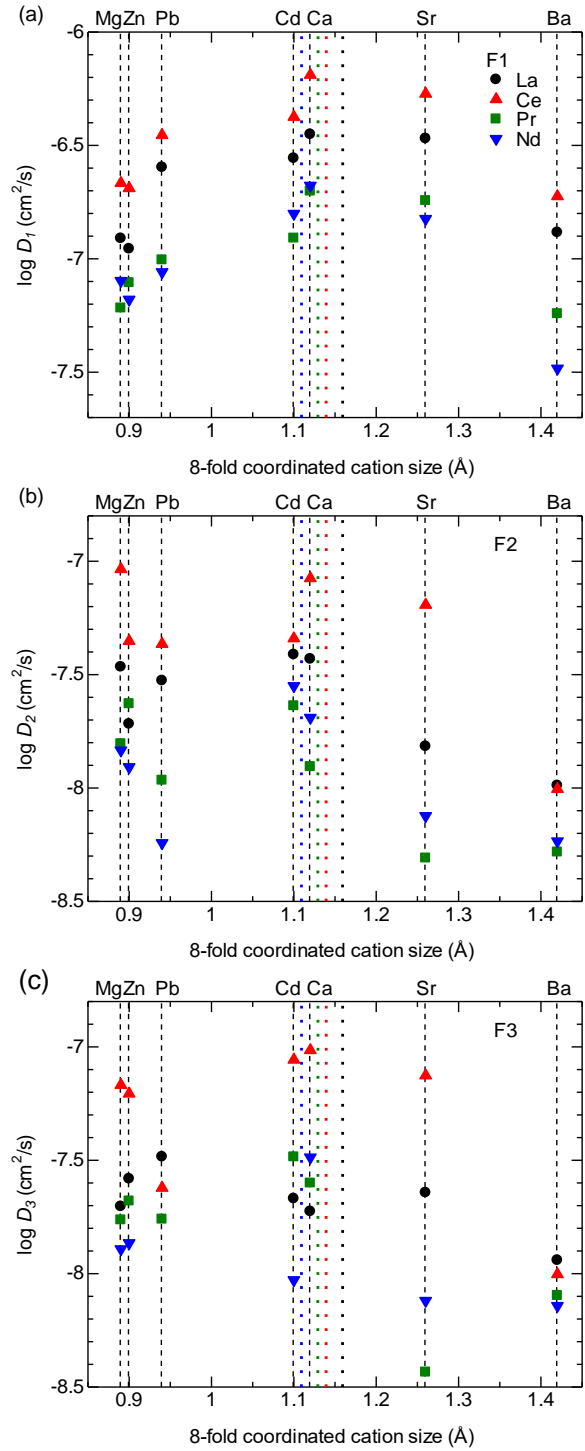

Fig. 2. Relation between (a) $D_1$, (b) $D_2$, and (c) $D_3$ for each trivalent cation versus the divalent cation size at $T$=500 K ($1/T$=0.002 K$^{-1}$). The eight-fold coordinated cation size is shown in vertical lines. Dashed lines with labels are used for divalent cations, while thick dotted lines in are used for trivalent cations (order of cation size is Nd$^{3+}$<Pr$^{3+}$<Ce$^{3+}$<La$^{3+}$).

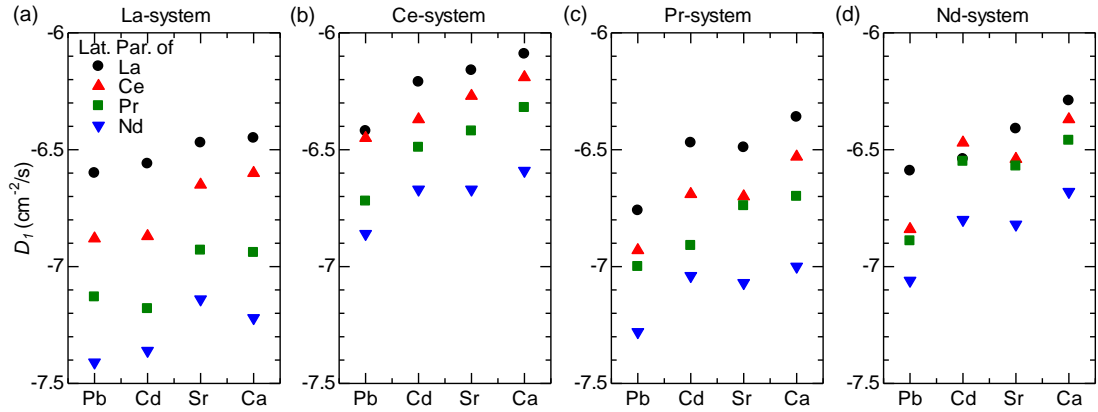

Fig. 3. $D_1$, for each divalent cation choice, when the lattice parameters are changed to $LaF_3$, $CeF_3$, $PrF_3$, and $NdF_3$ (black circles, red upward pointing triangles, green squares, and blue downward pointing triangles, respectively) when the trivalent cation is (a) La, (b) Ce, (c) Pr, and (d) Nd, respectively.

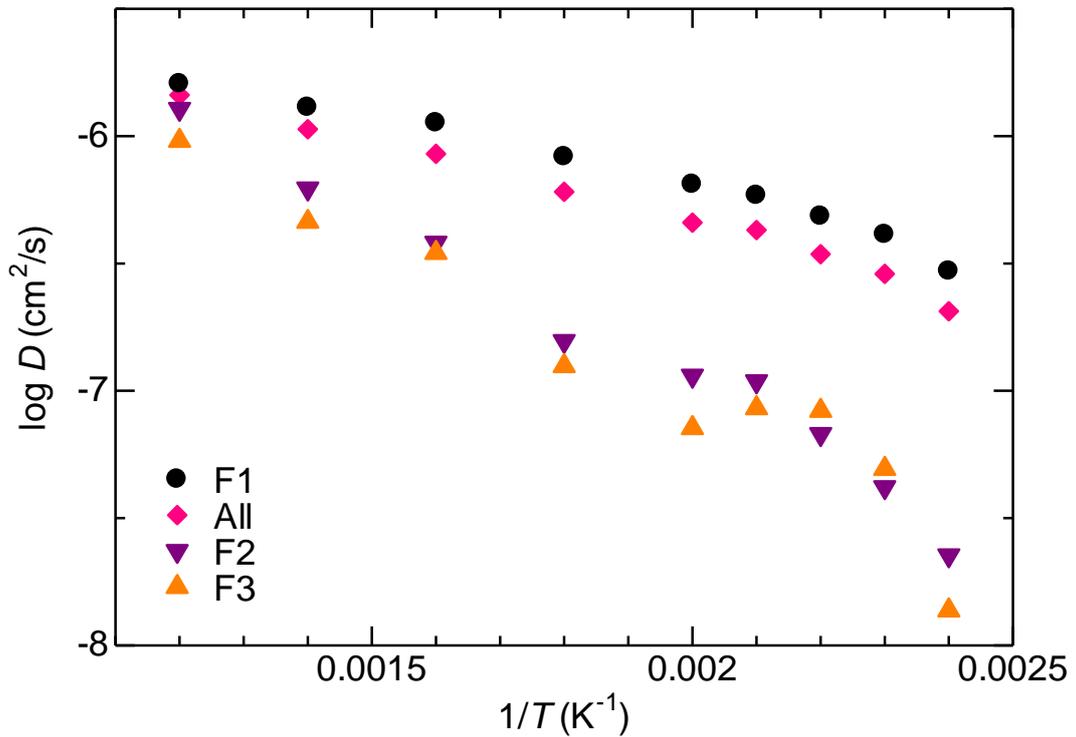

Fig. 4. Arrhenius plot of $Ce_{0.95}Ca_{0.05}F_{2.95}$ when the diffusion coefficient is averaged over different F types.

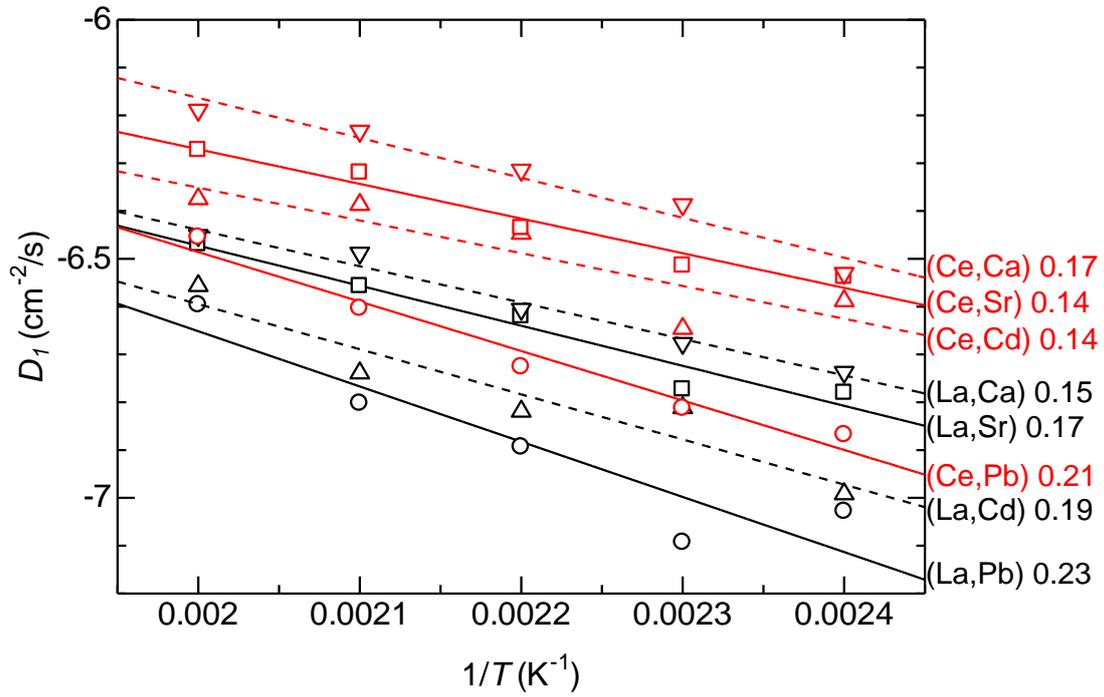

Fig. 5. Arrhenius plot of $D_1$ versus $1/T$ in $(La,Ce)_{0.95}(Pb,Cd,Sr,Ca)_{0.05}F_{2.95}$. The cation choices are shown for each plot. The symbol color is black and red for La and Ce, respectively, and shapes are empty downward pointing triangles, squares, upward pointing triangles, and circles for Ca, Sr, Cd, and Pb, respectively.

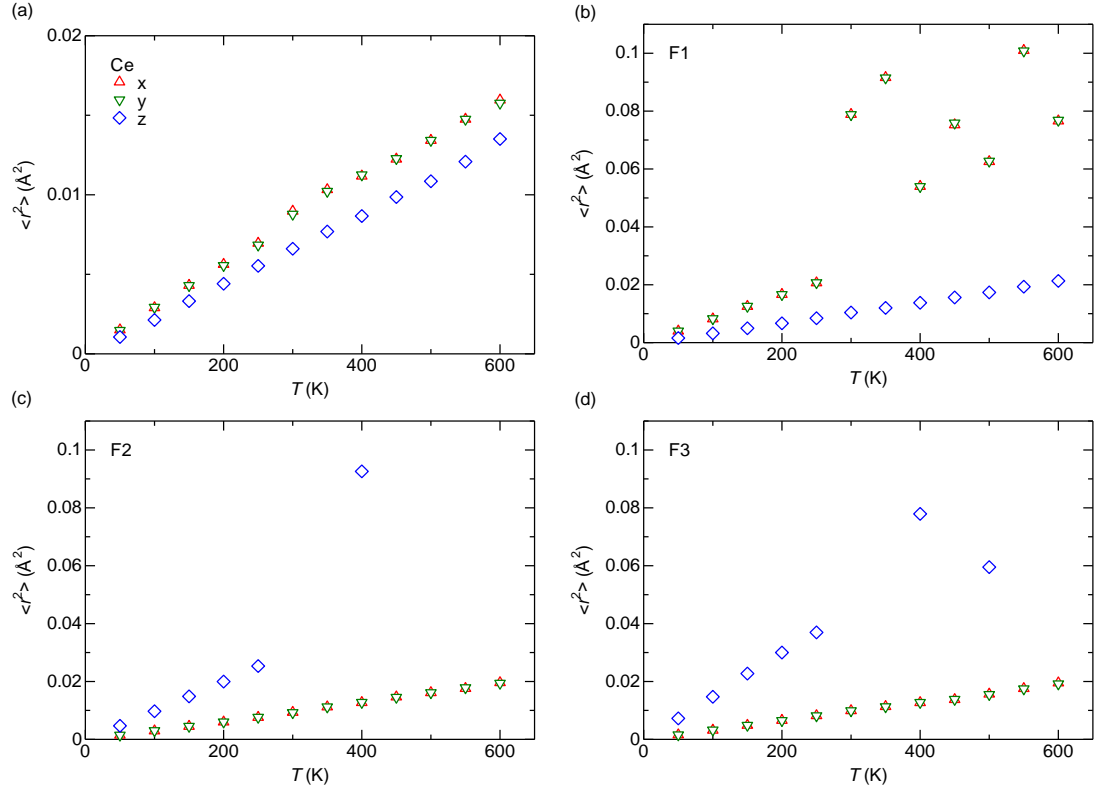

Fig. 6. <Δ$x^2$>, <Δ$y^2$>, and <Δ$z^2$> (red up-pointing triangles, green down-pointing triangles, and blue diamonds, respectively) of (a) Ce, (b) F1, (c) F2, and (d) F3 sites of CeF$_3$. Points for <Δ$x^2$> and <Δ$y^2$> almost overlap due to symmetry.

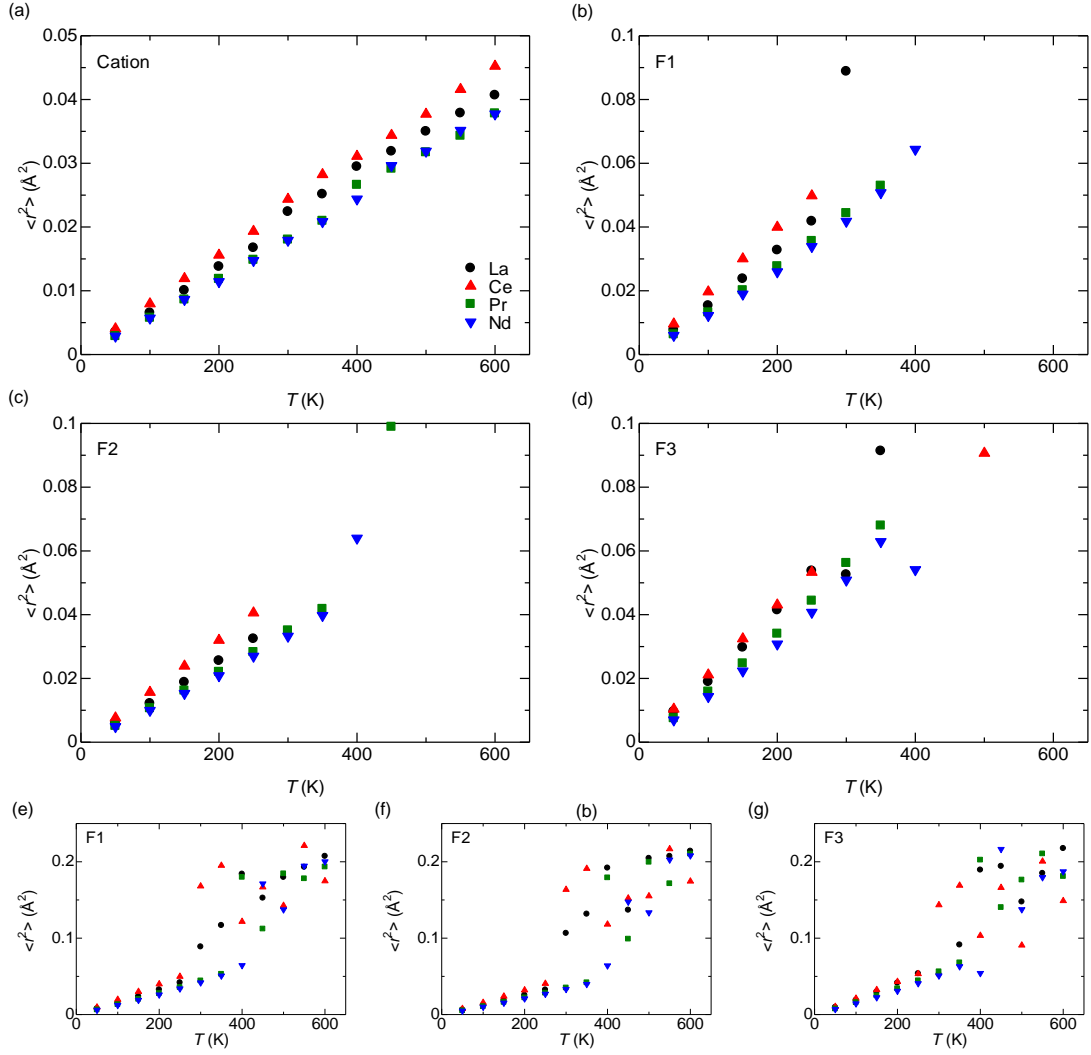

Fig. 7. $\langle \Delta r^2 \rangle = \langle \Delta x^2 \rangle + \langle \Delta y^2 \rangle + \langle \Delta z^2 \rangle$ for (a) cation sites, (b,e) F1 sites, (c,f) F2 sites, and (d,g) F3 sites of (La,Ce,Pr,Nd)F$_3$. (b-d) enlarges the low temperature region, while (e-g) show all points for F sites.

Graphical abstract

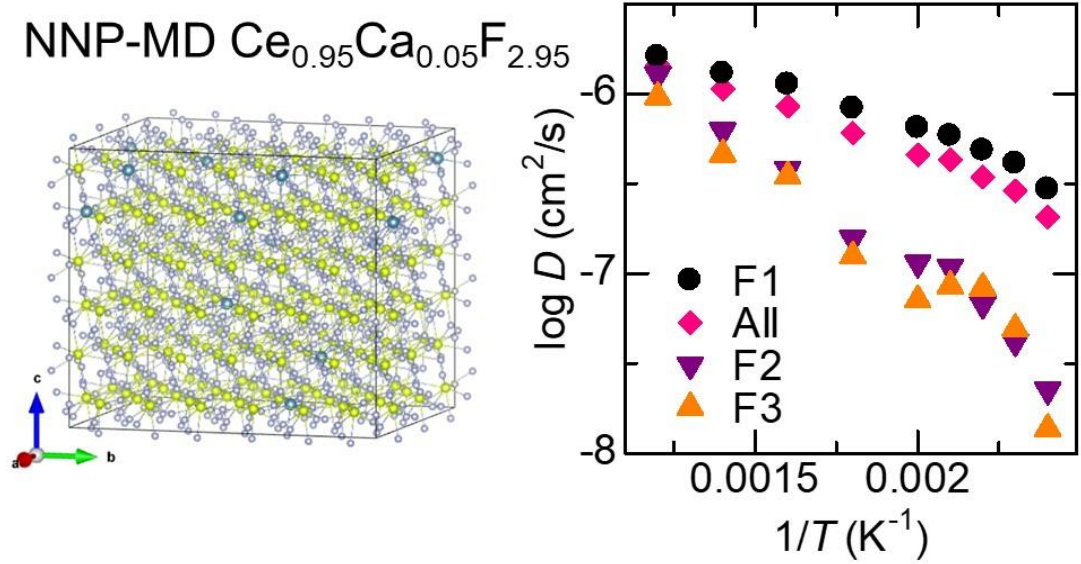